\documentclass{osa-article}

\journal{oe}
\articletype{Research Article}

\makeatletter
\renewcommand{\maketag@@@}[1]{\hbox{\m@th\normalsize\normalfont#1}}%
\makeatother

\begin{document}

\title{Observing two-photon subwavelength interference of broadband chaotic light in polarization-selective Michelson interferometer}

\author{Sheng Luo,\authormark{1,2} Yu Zhou,\authormark{1,4} Huaibin Zheng,\authormark{2,5} Wanting Xu,\authormark{3} Jianbin Liu,\authormark{2} Hui Chen,\authormark{2} Yuchen He,\authormark{2} Shuanghao Zhang,\authormark{2} Fuli Li,\authormark{1} and Zhuo Xu\authormark{2}}

\address{\authormark{1}MOE Key Laboratory for Nonequilibrium Synthesis and Modulation of Condensed Matter, Department of Applied Physics, Xi'an Jiaotong University, Xi'an 710049, China\\

\authormark{2}Electronic Materials Research Laboratory, Key Laboratory of the Ministry of Education $\&$ International Center for Dielectric Research, School of Electronic Science and Engineering, Xi'an Jiaotong University, Xi'an 710049, China\\

\authormark{3}School of Science, Xi'an Polytechnic University, Xi'an, Shaanxi 710048, China\\
\authormark{4}zhou1@mail.xjtu.edu.cn\\
\authormark{5}huaibinzheng@mail.xjtu.edu.cn} 


\begin{abstract}
Differing from the traditional method to achieve subwavelength interference, we have demonstrated the two-photon subwavelength interference effect of broadband chaotic light in polarization-selective Michelson interferometer with an ultrafast two-photon absorption detector for the first time, which is achieved by manipulating two-photon probability amplitudes involved in the interference. In theory, the two-photon polarization coherence matrix and probability amplitudes matrix are combined to develop polarized two-photon interference terms, which explains the experimental results well. In order to make better use of this interferometer to produce the subwavelength effect, we also make a series of error analyses to find out the relationship between the visibility and the degree of polarization error. Our experimental and theoretical results are helpful to understand the two-photon subwavelength interference, which sheds light on the development of the two-photon interference theory of vector light field based on quantum mechanics. These experimental results may help to develop future optical interferometry, optical polarimetry, and subwavelength lithography.
\end{abstract}

\section{Introduction}
How to break through Rayleigh limit has always attracted people's attention, mainly because it plays an important role in optical lithography and biological imaging\cite{lipson_lipson_lipson_2010,Introduction}. Jacobson et al. theoretically proposed the concept of Photonic de Broglie waves in 1995 and pointed out that the corresponding de Broglie wavelength is $\lambda/N$ for the compound system of $N$ photon entanglement, where $\lambda$ is the wavelength of a single-photon\cite{PhysRevLett.74.4835}. And then Fonseca et al. experimentally measured the de Broglie wavelength of two-photon entangled states which generated by type-II Spontaneous Parametric Down-Conversion (SPDC) for the first time\cite{PhysRevLett.82.2868}. Later, Boto and D 'Angelo et al. proposed quantum lithography that the resolution could beaten the classical diffraction limit by a factor of 2 via the two-photon entangled subwavelength interference\cite{PhysRevLett.85.2733,PhysRevLett.87.013602}. However, it is difficult to generate multi-photon entangled states using the nonlinear interaction between light and matter since the efficiency is low, which is not conducive to practical applications\cite{doi:10.1063/1.3472112,PhysRevA.88.043838,PhysRevA.91.053830}. Subsequently, theoretical and experimental studies have shown that thermal light play a similar role to the quantum entanglement of two-photon state in subwavelength interference\cite{PhysRevA.70.041801,Scarcelli_2004,PhysRevLett.94.173601,PhysRevA.72.043805}. Two-photon subwavelength interference of pseudothermal light has been measured in the Hanbury Brown-Twiss (HBT) interferometer when the two detectors are scanned in the opposite directions, whereas it also makes the high-order correlation subwavelength interference of thermal light can not be directly applied to quantum lithography\cite{brown1956correlation,brown1956test,PhysRevA.70.041801,Scarcelli_2004,PhysRevLett.94.173601,PhysRevA.72.043805,PhysRevA.82.013822}. Later, the methods of reversing the wave front and the design of new classical light sources or other ways have been proposed to overcome these defects\cite{PhysRevLett.96.163603,doi:10.1063/1.3472112,PhysRevA.88.043838,PhysRevA.91.053830,PhysRevA.87.023818}.

Up to now, the reported light sources that can achieve two-photon subwavelength interference including entangled photon pairs\cite{PhysRevLett.74.3600,PhysRevLett.74.4835,PhysRevLett.82.2868,PhysRevLett.85.2733,PhysRevLett.87.013602}, pseudothermal light\cite{PhysRevA.70.041801,Scarcelli_2004,PhysRevLett.94.173601,PhysRevA.72.043805}, coherent light and so on\cite{PhysRevLett.96.163603,PhysRevA.87.023818,Bentley:04,Peer:04,PhysRevA.97.043807}. As far as we know, the two-photon subwavelength interference effect of true broadband chaotic light has never been observed, mainly because the two-photon subwavelength interference effect of chaotic light is difficult to detect due to its femtosecond scale coherence time. It was not until 2009 that Boitier et al. proposed to use a detector based on two-photon absorption (TPA) to measure the photon bunching effect\cite{boitier2009measuring}. The detection response rate makes it can obtain the signal on the order of a few femtosecond\cite{Hayat2011Applications,nevet2011ultrafast}. In this way, many meaningful phenomena have measured based on TPA detection, such as the extrabunching effect\cite{Boitier:10,boitier2011photon}, ultrabroadband ghost imaging\cite{hartmann2015ultrabroadband}, ghost polarimetry\cite{Shevchenko:15,Janassek:18,shevchenko2017polarization,PhysRevResearch.2.012053}, and superbunching effect of broadband chaotic light\cite{PhysRevA.103.013723}. In fact, the counts detected by the TPA detector contains many different two-photon interference information, including the second-order interference term, the background term, the subwavelength interference term and so on. Each of these terms contains corresponding information about the interference and is worth studying. Based on this, we hope to use a way to control the two-photon probability amplitudes terms to modulate the two-photon subwavelength interference phenomenon based on TPA detection.

In this paper, we proposed and demonstrated a method to achieve temporal subwavelength effect of true broadband chaotic light in polarization-selective Michelson interferometer based on TPA detection. Also, we developed the polarized two-photon interference matrix by introducing the two-photon polarization coherence matrix in the quantum two-photon interference theory, which is in good agreement with the experimental results. It is different from the traditional method to achieve subwavelength interference, our technique can allows us to measure the two-photon subwavelength interference of chaotic stationary light at the timescale of a few femtoseconds, which may find potential applications in the quantum lithography, optical interferometer and optical polarimeter.

The remaining parts of the paper are organized as follows. In Sec. \ref{theory}, we will combine the quantum two-photon interference theory with the polarization theory to interpret the observed two-photon subwavelength interference. In Sec. \ref{experiment}, we will manipulate the two-photon interference phenomena by placing three polarizers at different positions in the Michelson interferometer. The discussions about the physics of the subwavelength interference and its experimental error analysis are in Sec. \ref{disccusion}. The conclusion of the paper is in Sec. \ref{conclusion}.

\section{THEORY}\label{theory}
Generally, when the quantum two-photon interference theory was employed to study the second-order coherence effect of thermal light, the usual practice was to simplify the light field as scalar field\cite{zhou2017superbunching,mandel1995optical,PhysRevA.103.013723}, and ignore the polarization characteristics of light produced by the vectorial electromagnetic field. Once polarizers are inserted in the optical path, the input light will change from unpolarized light to polarized light. Then the light we study will change from scalar to vector, so that the corresponding two-photon interference terms will change accordingly and result in different interference phenomena\cite{zhou1}, such as the two-photon subwavelength interference effect.

It is well known that the bunching effect of chaotic stationary light in the Michelson interferometer can be explained by quantum two-photon interference theory\cite{8526287,PhysRevA.103.013723}. When we develop vector two-photon interference theory, polarization is treat as a new dimension in the phase space of system, just like spatial, temporal, and spectral coherence properties. Coherent superposition of two-photon probability amplitudes happens only in the same coherent mode of polarization also like those happen in spatial and temporal domain\cite{loudon1973the,PhysRevA.72.043805,zhou2017superbunching}. On the other hand, the polarization can act as a switch to manipulate the quantum two-photon interference happens in this selective Michelson interferometer. Many new effect can be observed by manipulating polarization setups and subwavelength effect is one of them. In order to develop the theory of two-photon subwavelength interference, we need to matrix the two-photon interference terms and then bring the two-photon polarization coherence matrix into the related TPA detection probability amplitudes matrix, and finally achieve the two-photon subwavelength effect\cite{zhou1}. The two-photon interference matrix based on TPA detection modulated by arbitrary polarizers can be expressed as
\begin{equation}\label{equ1}
{{\bf{M}}_{\bf{P}}} = {{\bf{J}}_{\bf{P}}}\cdot{{\bf{M}}_{\bf{TPA}}},
\end{equation}
where ${{\bf{M}}_{\bf{TPA}}}$ is the two-photon probability amplitudes matrix, ${{\bf{J}}_{\bf{P}}}$ is the polarization coefficient matrix of the two-photon coherence terms. Also ${{\bf{J}}_{\bf{P}}}$ determines whether the TPA detection probability amplitudes terms are present, just like a switch.

Firstly, we briefly review the quantum two-photon interference theory based on Michelson interferometer. The two-photon bunching effect is caused by the superposition of the probability amplitudes corresponding to different and indistinguishable paths triggering the TPA detector. We define $a$ and $b$ as two photons from the chaotic light. The channel 1 and channel 2 can be defined as path that the photons are reflected from $\rm M_1$ and $\rm M_2$, respectively. There are four different but indistinguishable ways for photons $a$ and $b$ to trigger the TPA detector, corresponding to four two-photon probability amplitudes which are $A_{a 1 b 1}, A_{a 1 b 2}, A_{a 2 b 1}$ and $A_{a 2 b 2}$. The probability of a TPA detection event happening is
\begin{equation}\label{equ2}
\begin{aligned}
C_{TPA}&=\left\langle|A_{a 1 b 1}+A_{a 1 b 2}+A_{a 2 b 1}+A_{a2b2}|^{2}\right\rangle\\
&=\left\langle {\left( \begin{array}{l}
{A_{a1b1}}A_{a1b1}^{\rm{*}} + {A_{a1b1}}A_{a1b2}^{\rm{*}}{\rm{ + }}{A_{a1b1}}A_{a2b1}^{\rm{*}}{\rm{ + }}{A_{a1b1}}A_{a2b2}^{\rm{*}}{\rm{ + }}\\
{A_{a1b2}}A_{a1b1}^{\rm{*}} + {A_{a1b2}}A_{a1b2}^{\rm{*}}{\rm{ + }}{A_{a1b2}}A_{a2b1}^{\rm{*}}{\rm{ + }}{A_{a1b2}}A_{a2b2}^{\rm{*}}{\rm{ + }}\\
{A_{a2b1}}A_{a1b1}^{\rm{*}} + {A_{a2b1}}A_{a1b2}^{\rm{*}}{\rm{ + }}{A_{a2b1}}A_{a2b1}^{\rm{*}}{\rm{ + }}{A_{a2b1}}A_{a2b2}^{\rm{*}}{\rm{ + }}\\
{A_{a2b2}}A_{a1b1}^{\rm{*}}{\rm{ + }}{A_{a2b2}}A_{a1b2}^{\rm{*}}{\rm{ + }}{A_{a2b2}}A_{a2b1}^{\rm{*}}{\rm{ + }}{A_{a2b2}}A_{a2b2}^{\rm{*}}
\end{array} \right)} \right\rangle
\end{aligned},
\end{equation}
where ${\left\langle {\cdot \cdot \cdot } \right\rangle}$ is the ensemble average by taking all the possible realizations into account. The two-photon probability amplitede $A_{aibj}$ indicates that photon $a$ is reflected by mirror $i$ (in arm $i$) and photon $b$ is refected by mirror $j$ (in arm $j$) before triggering the TPA detector together. Where $A_{aibj}^{*}$ is the complex conjugate of $A_{aibj}$, $i,j=1,2$.

There are 16 probability amplitudes multiplied by each other, which we can view as the sum of all the terms in a $4\times4$ matrix. Therefore, the probability amplitudes matrix based on TPA detection can be expressed as

\begin{equation}\label{equ3}
\begin{aligned}
{{\bf{M}}_{\bf{TPA}}} = \left( {\begin{array}{*{20}{c}}
{{A_{a1b1}}A_{a1b1}^{\rm{*}}}&{{A_{a1b1}}A_{a1b2}^{\rm{*}}}&{{A_{a1b1}}A_{a2b1}^{\rm{*}}}&{{A_{a1b1}}A_{a2b2}^{\rm{*}}}\\
{{A_{a1b2}}A_{a1b1}^{\rm{*}}}&{{A_{a1b2}}A_{a1b2}^{\rm{*}}}&{{A_{a1b2}}A_{a2b1}^{\rm{*}}}&{{A_{a1b2}}A_{a2b2}^{\rm{*}}}\\
{{A_{a2b1}}A_{a1b1}^{\rm{*}}}&{{A_{a2b1}}A_{a1b2}^{\rm{*}}}&{{A_{a2b1}}A_{a2b1}^{\rm{*}}}&{{A_{a2b1}}A_{a2b2}^{\rm{*}}}\\
{{A_{a2b2}}A_{a1b1}^{\rm{*}}}&{{A_{a2b2}}A_{a1b2}^{\rm{*}}}&{{A_{a2b2}}A_{a2b1}^{\rm{*}}}&{{A_{a2b2}}A_{a2b2}^{\rm{*}}}
\end{array}} \right)
\end{aligned}.
\end{equation}

When we add three polarizers with different polarization angles to the optical path, where $\rm P_1$ is at the input of the interferometer and set to $0^{\circ}$ polarization (horizontal), $\rm P_2$ and $\rm P_3$ are set to $45^{\circ}$ with respect to the horizontal polarization at the two arms of interferometer, the light we study will change from scalar to vector according to the Refs\cite{zhou1,mandel1995optical}. The polarization coefficient matrix of the two-photon coherence terms can be expressed as
\begin{equation}\label{equ4}
{{\bf{J}}_{\bf{P}}}{\rm{ = }}\left( {\begin{array}{*{20}{c}}
{\frac{3}{8}}&0&0&{ - \frac{1}{8}}\\
0&{\frac{1}{8}}&{\frac{1}{8}}&0\\
0&{\frac{1}{8}}&{\frac{1}{8}}&0\\
{ - \frac{1}{8}}&0&0&{\frac{3}{8}}
\end{array}} \right).
\end{equation}

\begin{figure}[htb]
\centering\includegraphics[scale=0.3]{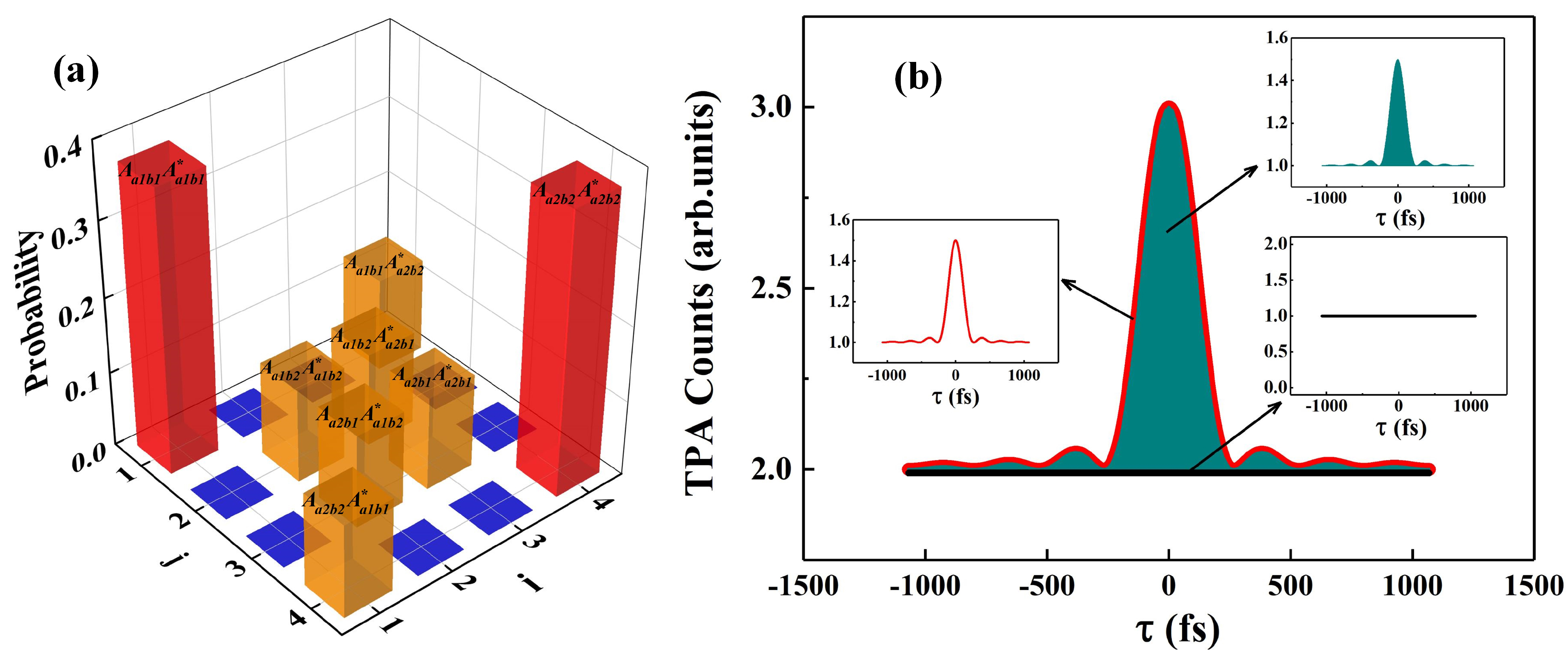}
\caption{\label{1} (a) shows the matrix visualization of the absolute value of ${{\bf{J}}_{{\bf{P}}}}$, the values represented by red, yellow and blue histogram are $\frac{3}{8}$, $\frac{1}{8}$ and $0$ respectively. The probability amplitudes above the histogram represent the two-photon interference terms in Eq.~(\ref{equ3}) can be modulated by this polarization matrix. (b) shows the simulation results of two-photon interferogram with three polarizers, which corresponds to Eq.~(\ref{equ6}).}
\end{figure}

For the sake of showing which two-photon probability amplitudes terms are modulated by the two-photon polarization coherence matrix more directly, we take the absolute value of ${{\bf{J}}_{{\bf{P}}}}$ and its matrix visualization is shown in Fig. \ref{1}(a). $i$, $j$ represent the columns and rows of the matrix, respectively.

According to the Feynman path integration theory \cite{book1,PhysRevA.103.013723}, the probability amplitude can be expressed as $A_{aibj}=e^{i \varphi_{a}} K_{a i} e^{i \varphi_{b}} K_{b j}$, where $e^{i \varphi_{a}}$ and $e^{i \varphi_{b}}$ are the initial phase of photon $a$ and photon $b$, respectively. $K_{a i}$, $K_{b j}$ are the Feynman propagators of different photons. Since the thermal light source has a certain spectral distribution, which is different from the single frequency of the laser. In this case, the Feynman propagator of the temporal correlation from the thermal light can be expressed as $K\left(t_{1}-t_{2}\right)=\int_{\omega_{0}-\frac{1}{2} \Delta \omega}^{\omega_{0}+\frac{1}{2} \Delta \omega}f(\omega)e^{-i\omega\left(t_{1}-t_{2}\right)} d \omega$, where $f(\omega)$ is the spectral distribution function of light source that assumed the rectangular spectrum distribution, $\omega_{0}$ is the center frequency and the spectral bandwidth is $\Delta\omega$, which all parameters depend on the spectral characteristics of the light source we are using. $t_{1}-t_{2}$ is the time difference that equals $\tau$. Substitute Eq.~(\ref{equ3}) and Eq.~(\ref{equ4}) into Eq.~(\ref{equ1}), the polarized two-photon interference matrix can be simplified as
\begin{footnotesize}
\begin{equation}\label{equ5}
{{\bf{M}}_{{\bf{P}}}}{\rm{ = }}\left( {\begin{array}{*{20}{c}}
{\frac{3}{8}}&0&0&{{\rm{ - }}\frac{1}{8}{{\rm{e}}^{ - 2i\tau {\omega _0}}}sin{c^2}\left( {\frac{{\Delta \omega \tau }}{2}} \right)}\\
0&{\frac{1}{8}}&{\frac{1}{8}sin{c^2}\left( {\frac{{\Delta \omega \tau }}{2}} \right)}&0\\
0&{\frac{1}{8}sin{c^2}\left( {\frac{{\Delta \omega \tau }}{2}} \right)}&{\frac{1}{8}}&0\\
{{\rm{ - }}\frac{1}{8}{{\rm{e}}^{2i\tau {\omega _0}}}sin{c^2}\left( {\frac{{\Delta \omega \tau }}{2}} \right)}&0&0&{\frac{3}{8}}
\end{array}} \right)
\end{equation}
\end{footnotesize}

After adding each term of the two-photon interference matrix ${{\bf{M}}_{{\bf{P}}}}$, we can get the normalized two-photon interference phenomenon based on TPA detection controlled by polarizers is
\begin{equation}\label{equ6}
\begin{array}{c}
{C_P}= 2 + \frac{1}{2}sin{c^2}\left( {\frac{1}{2}\Delta \omega \tau } \right) + \frac{1}{2}\cos \left( {2{\omega _0}\tau } \right)sin{c^2}\left( {\frac{1}{2}\Delta \omega \tau } \right)
\end{array}
\end{equation}

The simulation results of Eq.~(\ref{equ6}) are shown in Fig. \ref{1}(b), it is actually made up of three different two-photon interference terms. $1{\rm{ + }}sin{c^2}\left( {\frac{1}{2}\Delta \omega \tau } \right)$ is the second-order coherence function corresponding to the solid red line in the illustration.
$1{\rm{ + }}\cos \left( {2{\omega _0}\tau } \right)sin{c^2}\left( {\frac{1}{2}\Delta \omega \tau } \right)$ is the two-photon subwavelength interference term corresponding to the solid green line in the illustration, where its oscillation frequency is $2{\omega _0}$, which means that the measured wavelength will be half the original wavelength. 1 is the constant term corresponding to the solid black line in the illustration.

\section{EXPERIMENT}\label{experiment}
\begin{figure}[htb]
\centering\includegraphics[scale=0.45]{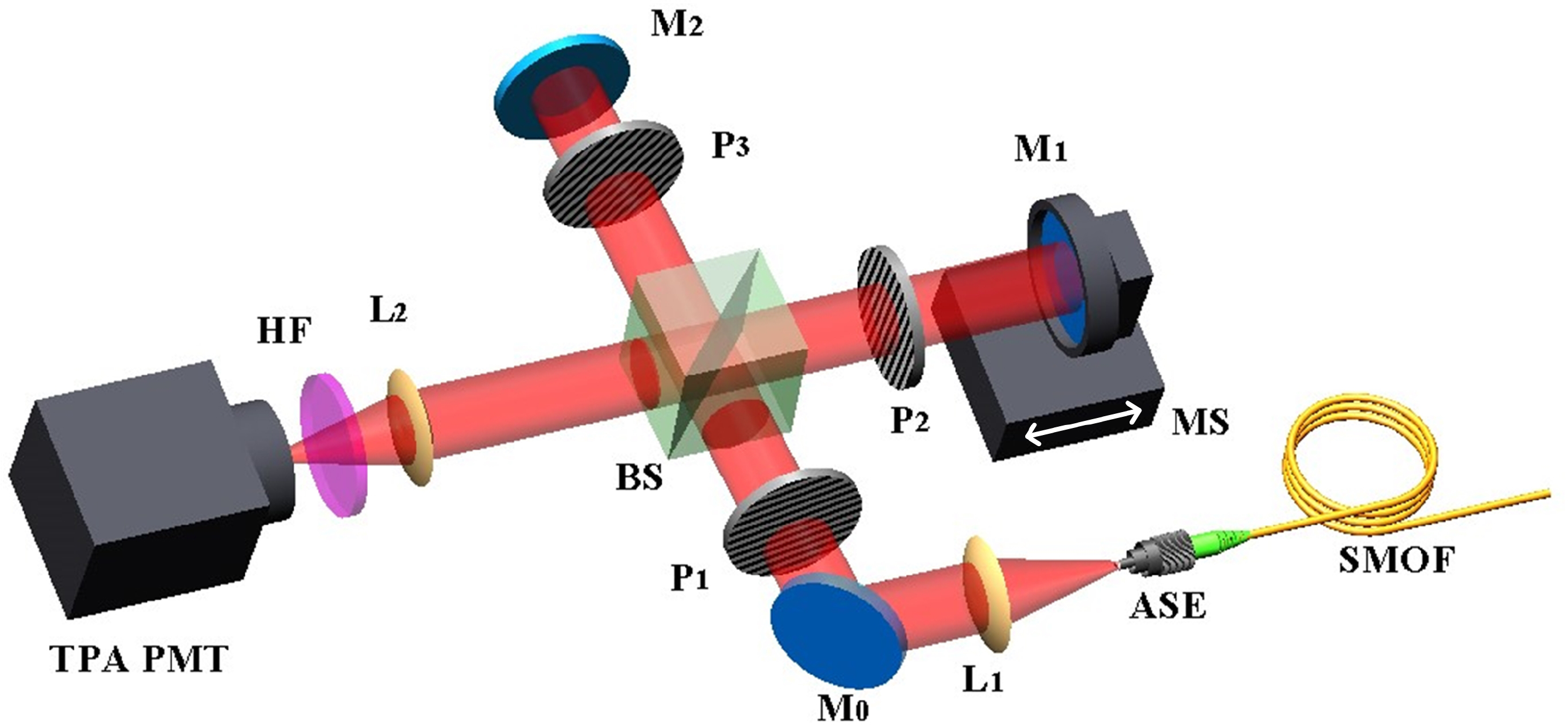}
\caption{\label{2} Experimental setup used to modulate the two-photon subwavelength interference based on the polarized Michelson interferometer. $\rm M_1$ and $\rm M_2$ are 1550 nm dielectric mirrors, where $\rm M_1$ is fixed on a precise motorized linear translation stage (MS). $\rm P_1$, $\rm P_2$, $\rm P_3$ are linear polarizers. SMOF is single-mode optical fiber. PMT is a GaAs photomultiplier tube.}
\end{figure}

By placing three polarizers at different positions in the Michelson interferometer, we can observe two-photon subwavelength interference phenomenon of broadband chaotic light. The experimental setup is shown schematically in Fig. \ref{2}. A continuous amplified spontaneous emission (ASE) incoherent light is used in the configuration, which is completely unpolarized light as like as natural light. This source serves as 1550 nm center wavelength whose bandwidth is 30 nm. Light that we analyze is coupled into a single-mode optical fiber (SMOF) and delivered to the Michelson interferometer which consists of two mirrors ($\rm M_1$ and $\rm M_2$) and a beam splitter (BS). These polarizers $\rm P_1$, $\rm P_2$, $\rm P_3$ will be placed as shown in the Fig. \ref{2}. $\rm P_1$ is at the input of the interferometer and set to $0^{\circ}$ polarization (horizontal). $\rm P_2$, $\rm P_3$ are set to $45^{\circ}$ with respect to the horizontal polarization, where are in front of $\rm M_1$, $\rm M_2$ respectively. The lens $\rm L_1$ and $\rm L_2$ are two convergent lenses with focal lengths of 10 mm and 25.4mm, respectively. The output beam of the interferometer is focused into a semiconductor photomultiplier tube (PMT) (Hamamatsu H7421-50) operated in two-photon absorption regime by $\rm L_2$. The high-pass filter (HF) with a cutoff wavelength of 1300 nm are used to eliminate the single-photon counting of the PMT. They are placed in front of the detector successively.

\begin{figure}[htb]
\centering\includegraphics[scale=0.57]{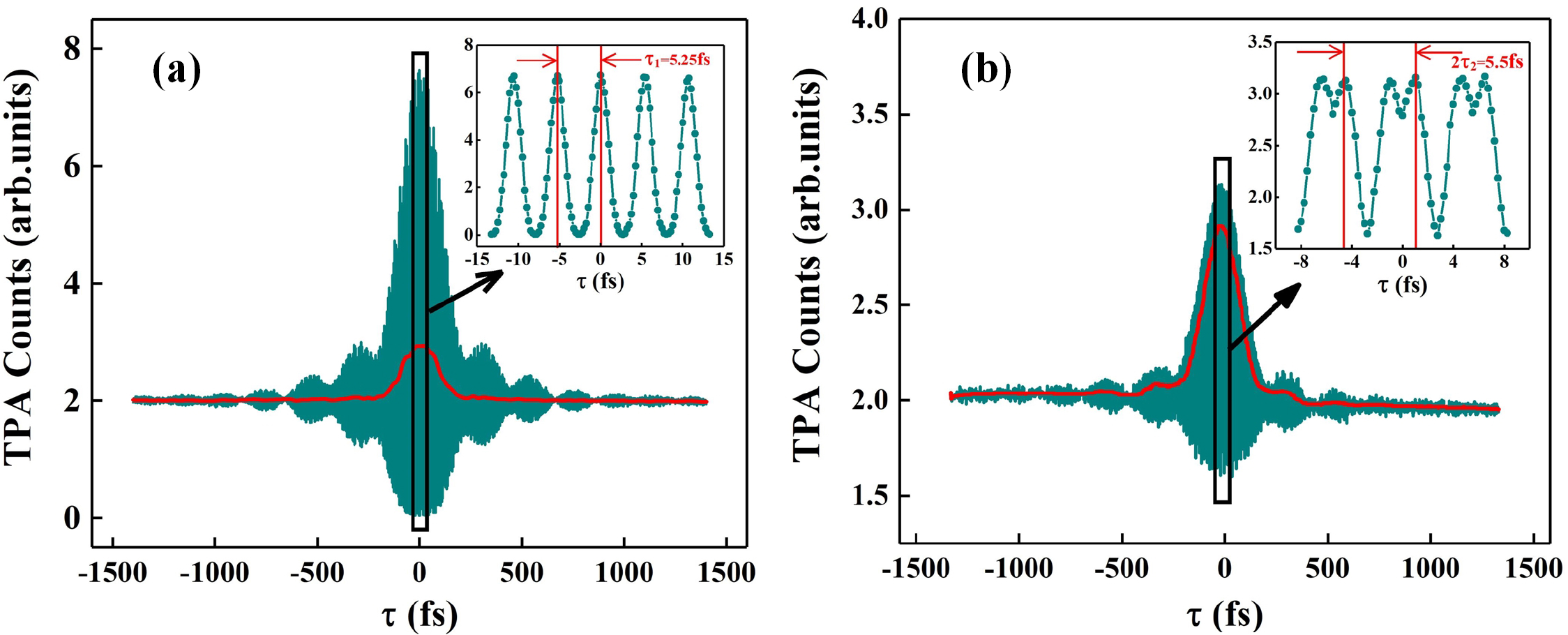}
\caption{\label{3} The measured two-photon absorption counts and the related intensity correlation functions under different conditions. (a) The measurement results of bunching effect in TPA based on that there is no polarizer at the interferometer. (b) The measurement results of two-photon subwavelength interference when
three polarizers ($\rm P_1$, $\rm P_2$, $\rm P_3$) with different polarization angles are placed in the Michelson interferometer as shown as in Fig. \ref{2}. The green line is a graph of all counting results of the TPA detector. The red lines represent the normalized second-order correlation function. The insets show a time-scale enlarged part in both interferograms.}
\end{figure}

In order to make a comparison between the ordinary bunching effect and two-photon subwavelength interference effect, we first measured its photon bunching effect. Firstly, there is no polarizers placing at the Michelson interferometer. The second-order coherence function of the bunching effect is measured by scanning one arm of the interferometer-scanning $\rm M_1$ on a motor stage (MS). The dark counts is 120 per second in the measurement. The TPA counts are recorded versus the time delay difference between two arms. The experimental results are shown in Fig. \ref{3}(a) by the green line, a fragment of which is shown in the inset for $\tau \in \left[ {{\rm{ - }}15,15} \right]$fs. The measured beating period is ${\tau _1} = 5.25$fs. The red solid line depicts the result of filtering out the high-frequency oscillation terms that is equal to the intensity correlation function obtained from the green curve, which corresponds to the degree of second-order coherence
${g^{(2)}}(0) = 1.92 \pm 0.02$ obtained after normalization. Since there is no polarizers, the measured bunching effect is caused by the interference of four two-photon probability amplitudes in the Eq.~(\ref{equ1}).

In the next step, $\rm P_1$, $\rm P_2$, $\rm P_3$ are inserted into the setup to constitute the polarization-selective Michelson interferometer as shown in Fig. \ref{2}. In this way, the ASE source will become horizontally linearly polarized light after passing through $\rm P_1$. Then the collimated beam from $\rm L_1$ is split at the BS, giving two beams which propagate through $\rm P_2$ and $\rm P_3$ in two different arms of the interferometer, where $\rm P_2$ and $\rm P_3$ are all set to $45^{\circ}$ with respect to the horizontal polarization in the clockwise direction. Although $\rm P_3$ is set to $45^{\circ}$, the beam reflected from $\rm M_2$ converged on BS and then reflected out of the Michelson interferometer, its polarization angle becomes $135^{\circ}$. Therefore, the two light beams from $\rm M_1$ and $\rm M_2$ are actually perpendicular to each other when they propagate through BS, which are $45^{\circ}$ and $135^{\circ}$ linearly polarized light respectively.

Two beams are combined at the output of the interferometer to trigger the TPA detector and the measurement results are shown in Fig. \ref{3}(b). The green line is the measured TPA counts by the polarization-selective Michelson interferometer. The red solid line is the result of removing the high-frequency oscillation term by numerical filtering, which corresponds to the degree of second-order coherence ${g^{(2)}}(0) = 1.89 \pm 0.01$. The most noteworthy phenomenon is that the beating period of the inserted detail of the measured interferogram is $2{\tau _2} = 5.5$fs and the average period is 2.75fs. This is almost half of the period measured in Fig. \ref{3}(a), which means that two-photon subwavelength interference is observed by selecting the two-photon probability amplitudes terms based on the polarization-selective Michelson interferometer.

In our experiment, there are some systematic errors, such as the placement of BS relative to $\rm M_3$, or the adjustment of the angle of the polarizers, it is extremely easy to cause the visibility of the two-photon subwavelength interference to decrease. The visibility of the two-photon subwavelength interference is only 6.2\%. Specific error analysis will be described in detail in Sec. \ref{disccusion}.

\section{DISCCUSION}\label{disccusion}
\begin{figure}[htb]
\centering\includegraphics[scale=0.33]{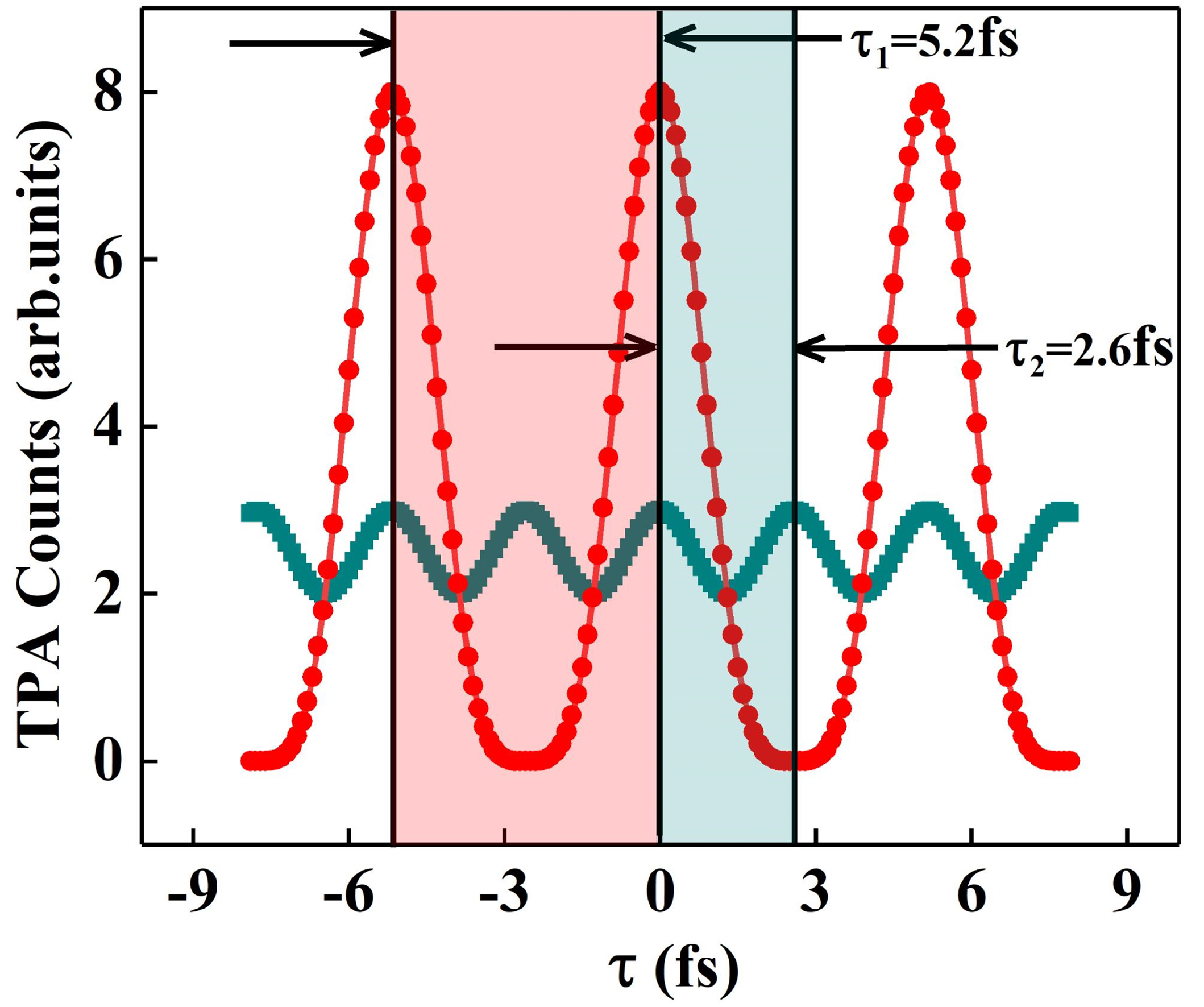}
\caption{\label{4} Simulation results of two-photon interference with different polarization conditions.
The red circles shows the details of Eq.~(\ref{equ2}) without polarizers. The green circles shows the details of Eq.~(\ref{equ6}) modulated by three polarizers, which corresponds to the two-photon subwavelength interference effect. The width of the red shadow and the green shadow represent their periods respectively.}
\end{figure}

To further study the two-photon subwavelength interference effect of broadband chaotic light, we simulate and compare the two-photon interference patterns based on TPA detection in two cases as shown in Fig. \ref{4}. The red circles represent the details of the TPA detection event when chaotic light transferred through the Michelson interferometer with no polarizers, which corresponds to shows the result of Eq.~(\ref{equ2}). The period of the interference pattern is 5.20 fs. The simulation also shows that the relative TPA counts will change from 8 to 0 when the phase changes from 0 to $\pi$. From these TPA counts a visibility $V = \left( {{C_{\max }} - {C_{\min }}} \right)/\left( {{C_{\max }} + {C_{\min }}} \right)$ with ${V} = 100\%$ is obtained. By contrast, the green circles represent the details of the TPA detection event based on polarization-selective Michelson interferometer, and it corresponds to the result of Eq.~(\ref{equ6}). It can be obviously found that the interference period of 2.60 fs is half of the previous two-photon interference pattern, and this is the two-photon subwavelength interference effect we observed. In addition, the visibility is reduced from 100\% to 20\%, and the TPA counts is reduced from 3 to 2 when $\tau$ changed by half a period.

\begin{figure}[htb]
\centering\includegraphics[scale=0.3]{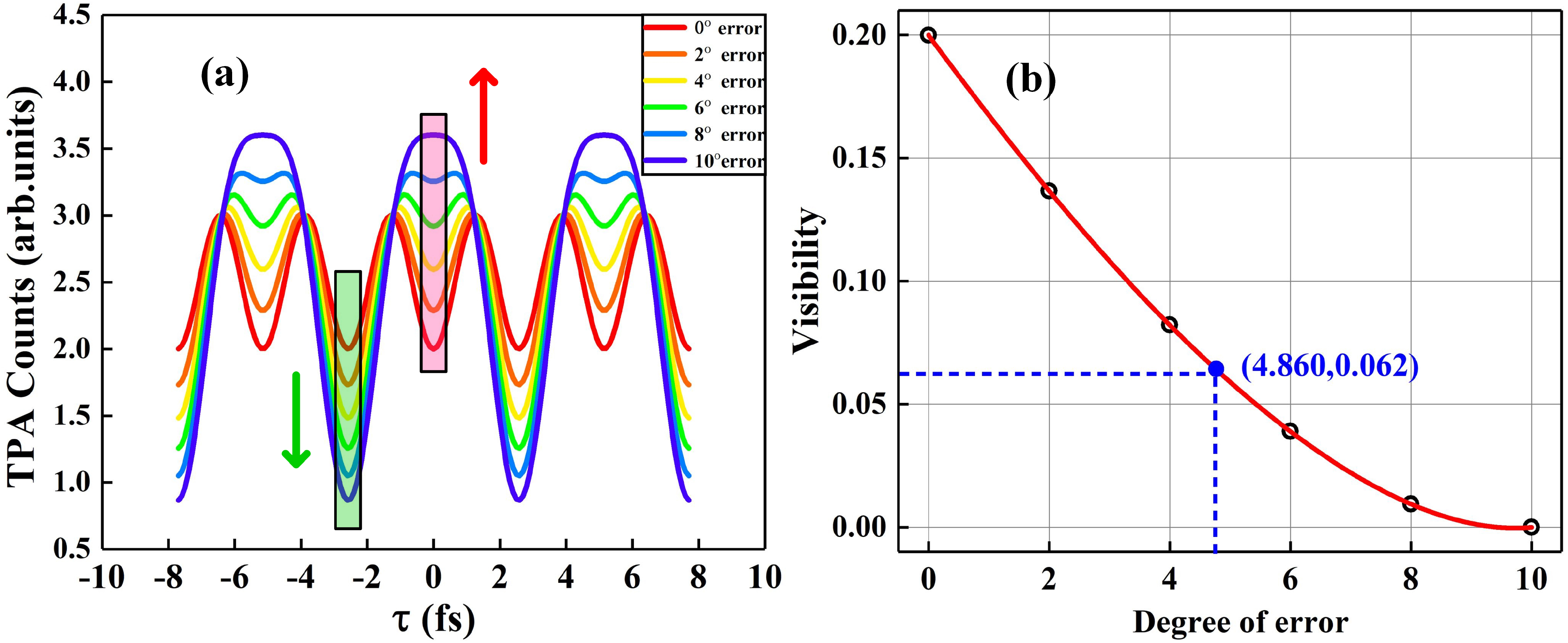}
\caption{\label{5} (a) shows the simulation results of two-photon subwavelength interference effect under different polarization errors. (b) shows the visibility of two-photon subwavelength interference effect under different polarization errors. The black circles are measured results and the red lines are fitted curve, the blue solid circle represents the polarization error and  visibility of two-photon subwavelength interference measured in Fig. \ref{3}(b).}
\end{figure}

Since there is a certain error between the measured two-photon subwavelength interference effect and the simulation results, we have done a set of error analysis under the same conditions to further study the source and size of the error in the two-photon subwavelength interference phenomenon. The simulation results of polarization degrees with different errors are shown in Fig. \ref{5}(a), the red solid line represents the simulation results in the most ideal experimental conditions with the polarization error of $0^{\circ}$. Its three polarizers ($\rm P_1$, $\rm P_2$, $\rm P_3$) are placed in the experimental setup as shown in Fig. \ref{2}, which the angle of polarization  are set to $0^{\circ}$, $45^{\circ}$ and $135^{\circ}$ respectively. Therefore, two-photon subwavelength interference effect can be observed, whose visibility reaches the maximum of 20\%.

However, it is obvious that the two-photon subwavelength interference effect is gradually disappearing, when we add a series of polarization errors into the ideal simulation system, such as $2^{\circ}$, $4^{\circ}$, $6^{\circ}$, $8^{\circ}$ and $10^{\circ}$. With the increase of the degree of polarization error, the counts of the red shadow in Fig. \ref{5}(a) will become higher and higher, which indicates that the visibility will decrease. While the green shadow of Fig. \ref{5}(a) will become lower and lower, indicating that more and more terms containing $\cos \left( {{\omega _0}\tau } \right)$ are mixed into Eq.~(\ref{equ6}). This is exactly the reason why the phenomenon of subwavelength interference effect measured in Fig. \ref{3}(b) has a lower background than the simulated background of 2 in Fig. \ref{1}(b). In addition, we also calculated the visibility of the two-photon subwavelength interference effect under different polarization errors as shown in Fig. \ref{5}(b). When the polarization error is from $0^{\circ}$ to $10^{\circ}$, the visibility drops from a maximum of 0.20 to 0. The black circles are measured results and the red line is the fitting of the measured data. Since the visibility of the two-photon subwavelength interference measured in experiment is 6.2\%, the polarization error is $4.860^{\circ} \pm 0.001^{\circ}$ according to the red fitted curve, as shown as the blue solid circles in the Fig. \ref{5}(b).

Therefore, it can be seen that the insertion of polarizers with specific angle in the TPA-based Michelson interference can indeed achieve the two-photon subwavelength interference effect, the essence is to introduce the two-photon polarization coherence matrix into the two-photon probability amplitudes matrix. However, in the actual experiment, we should pay special attention to the systematic error to avoid the error is too large to observe the two-photon subwavelength interference effect.

\section{CONCLUSION}\label{conclusion}
In summary, we have developed a polarization-selective Michelson interferometer to explore the two-photon subwavelength interference effect of broadband chaotic light based on TPA detection at a femtosecond timescale. The subwavelength effect is achieved by manipulating the quantum two-photon interference effect, in which we use polarizers as switches to control the coefficient of different two-photon probability amplitudes. By combining the two-photon polarization coherence matrix with TPA detection probability amplitudes matrix, the polarized two-photon interference matrix of the subwavelength interference effect is presented. The theoretical simulation results are in agreement with the experimental results. In order to better apply the two-photon subwavelength effect, we also make a series of error analysis to explain the visibility difference between the experimental results and the simulation results, and finally obtained that the polarization error of the system was $4.860^{\circ} \pm 0.001^{\circ}$.
The discussions of the two-photon subwavelength interference effect, in both the theory and experiment, are helpful to understand the second order coherence of the vector light field. We believe the results will be an important tool for studying the fundamental and applied aspects of optical coherence, such as the optical interferometry and subwavelength lithography, or other possible applications.

\section*{Funding}
This work was supported by Shaanxi Key Research and Development Project (Grant No. 2019ZDLGY09-09); National Natural Science Foundation of China (Grant No. 61901353); Key Innovation Team of Shaanxi Province (Grant No. 2018TD-024); 111 Project of China (Grant No. B14040).

\section*{Disclosures}

The authors declare no conflicts of interest.

\section*{Data availability}

Data underlying the results presented in this paper are not publicly available at this time but may be obtained from the authors upon reasonable request.


\bibliography{reference}

\end{document}